\documentclass[pra,aps,superscriptaddress,twocolumn,nopacs,nofootinbib, longbibliography]{revtex4-2}

\usepackage{array,multirow,graphicx,color,amsmath,amsfonts,enumerate,amsthm,amssymb,mathtools,enumitem,thmtools,hyperref,subfigure,mathdots,enumitem,centernot,bm,soul,bbm,tikz,pgfplots,soul}
\usepackage[capitalise, noabbrev]{cleveref}
\usepackage[utf8]{inputenc}
\usetikzlibrary{arrows}
\pgfplotsset{compat=1.14}
\hypersetup{colorlinks=true,linkcolor=blue,citecolor=blue,urlcolor=blue}

\def\S{ {\mathcal S} }
\def\F{ {\mathcal F} }

\def\F{ {\cal F} }

\def\>{\rangle}
\def\<{\langle}

\newcommand{\ket}[1]{| {#1} \rangle}

\definecolor{ppblue}{RGB}{46,117,182}
\definecolor{ppred}{RGB}{197, 90, 17}

\theoremstyle{plain}

\theoremstyle{definition}

\newcolumntype{C}[1]{>{\centering\arraybackslash}p{#1}}

\definecolor{tikzBlue}{rgb}{0.6941176470588235,0.7568627450980392,0.8588235294117647}
\definecolor{tikzOrange}{rgb}{0.9294117647058824,0.7647058823529411,0.49019607843137253}
\definecolor{tikzBlue2}{rgb}{0.462745098,0.504575163,0.57254902}
\definecolor{tikzOrange2}{rgb}{0.619607843,0.509803922,0.326797386}
\definecolor{tikzGray}{rgb}{0.7529411764705882,0.7529411764705882,0.7529411764705882}

\DeclareFontFamily{U}{mathb}{\hyphenchar\font45}
\DeclareFontShape{U}{mathb}{m}{n}{
	<-6> mathb5 <6-7> mathb6 <7-8> mathb7
	<8-9> mathb8 <9-10> mathb9
	<10-12> mathb10 <12-> mathb12
}{}
\DeclareSymbolFont{mathb}{U}{mathb}{m}{n}
\DeclareMathSymbol{\llcurly}{\mathrel}{mathb}{"CE}
\DeclareMathSymbol{\ggcurly}{\mathrel}{mathb}{"CF}

\begin{document}

\title{The original Wigner's friend paradox within a realist toy model}
\author{Matteo Lostaglio}
\affiliation{Korteweg-de Vries Institute for Mathematics and QuSoft, University of Amsterdam, The Netherlands}
\author{Joseph Bowles}
\affiliation{ICFO - Institut de Ci\`encies Fot\`oniques, The Barcelona Institute of Science and Technology, 08860 Castelldefels, Spain}
\begin{abstract}
The original Wigner's friend paradox is a gedankenexperiment involving an observer described by an external agent. The paradox highlights the tension between unitary evolution and collapse in quantum theory, and is sometimes taken as requiring a reassessment of the notion of objective reality. In this note however we present a classical toy model in which (i) The contradicting predictions at the heart of the thought experiment are reproduced (ii) Every system is in a well-defined state at all times. The toy model shows how  puzzles such as Wigner's friend's experience of being in a superposition, conflicts between different agents' descriptions of the experiment, the positioning of the Heisenberg's cut and the apparent lack of objectivity of measurement outcomes can be explained within a classical model where there exists an objective state of affairs about every physical system at all times. Within the model, the debate surrounding the original Wigner's friend thought experiment and its resolution have striking similarities with arguments concerning the nature of the second law of thermodynamics. The same conclusion however does not apply to more recent extensions of the gedankenexperiment featuring multiple encapsulated observers, and shows that such extensions are indeed necessary avoid simple classical explanations. 
\end{abstract}

\maketitle


\section{Setting the stage}

The original Wigner's friend gedankenexperiment is a thought experiment first considered by Wigner in 1961~\cite{wigner1961} that highlights the tension between unitary evolution and measurement collapse in quantum theory. It involves two agents, namely Wigner~($W$) and his friend~($F$).

In the simplest setting, a single qubit system is initially prepared in the state $\ket{0}_S$. $S$ undergoes a (Hadamard) unitary evolution, after which its state is $\frac{1}{\sqrt{2}} (\ket{0}_S + \ket{1}_S)$. At this point $S$ is measured by Wigner's friend, who finds it in state $\ket{0}_S$ or $\ket{1}_S$ with equal probability according to the Born rule.

Wigner is a superobserver sitting outside his friend's lab. According to him, the evolution of both $S$ and his friend $F$ can be described by Schr\"odinger's equation. In the most bare-bone description, $W$ describes the relevant degrees of freedom of $F$ as being themselves a two-level system, schematically representing a memory on which the measurement outcome is imprinted. The friend's memory state is initialised in $\ket{0}_F$ and the measurement of $S$ by $F$ can be described as a CNOT between the two:
\small
\begin{equation*}
	\frac{1}{\sqrt{2}} (\ket{0}_S + \ket{1}_S) \otimes \ket{0}_F \mapsto \frac{1}{\sqrt{2}} (\ket{0}_S \otimes \ket{0}_F  + \ket{1}_S \otimes \ket{1}_F ):= \ket{\phi^+}_{SF}.
\end{equation*}  
\normalsize
After the interaction, the states $\ket{0}_F$ and $\ket{1}_F$ correspond to the labels ``the friend sees outcome $0$'' and ``the friend sees outcome $1$''. Wigner can verify his prediction by performing a measurement in the basis of $2$-qubit maximally entangled states $\{ \ket{\psi^{\pm}}_{SF}, \ket{\phi^\pm}_{SF}\}$. He will get the outcome $\phi^+$ with probability $1$. Hence, if quantum theory is correct at all scales, the thought experiment requires that Wigner's description of the global entangled state must be compatible with Wigner's friend experience of a definite measurement outcome.

Note however that if $F$ applies the collapse rule she would assign to $S$ the state $\ket{0}_S$ \emph{or} $\ket{1}_S$. According to this procedure she would then predict that the two outcomes $\ket{\phi^\pm}_{SF}$ of Wigner's measurement can occur, which is in contradiction with $W$'s prediction of obtaining $\phi^+$ with probability $1$. Since we assume here and throughout the universality of quantum mechanics, this inference from the collapse rule must be unwarranted. 

Available explanations of this thought experiment compatible with quantum mechanics involve serious departures from the classical worldview. Broadly speaking they are of two kinds: perspectival interpretations such as QBism~\cite{fuchs2019qbism}, which  are at ease with the idea that every fact, including a measurement outcome, is relative to a particular observer; and interpretations such as Bohmian mechanics~\cite{durr2012quantum} or the many-world interpretation~\cite{dewitt2015many}, in which the quantum state is taken to be part of a highly nonclassical ontology and there is no actual collapse, which is understood as an effective procedure. So it is natural to ask whether classical models reproducing the same phenomenology exist, and how the abovementioned issues look like within such descriptions.

\begin{figure*}
	\centering
	\includegraphics[scale=0.6]{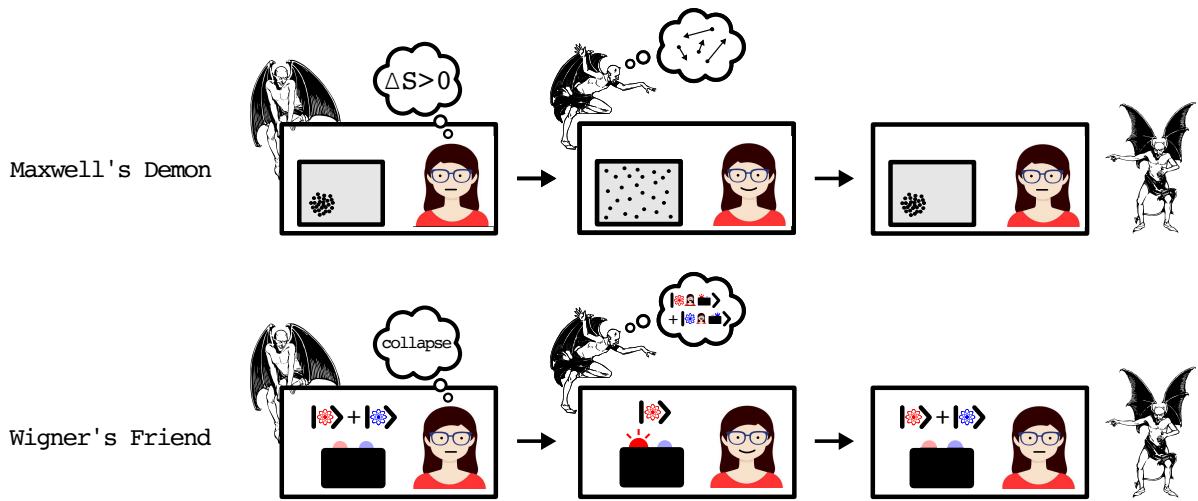}
	\caption{\emph{Top:} A physicist with limited control of a gas in a box predicts and observes an irreversible (to her) increase of entropy in a spontaneous equilibration process. A cheeky demon with full knowledge of positions and momenta of all particles can reverse the process bringing the gas (and in principle her too) back to the original state. \emph{Bottom:} Wigner's friend predicts and observes an irreversible (to her) collapse of the quantum state during a measurement process. A cheeky demon named Wigner with full knowledge of the global quantum state can reverse the process bringing system, measuring device and her back to the original state. In this note we highlight and discuss the analogies between these two setups and their significance for the debate surrounding the original Wigner's friend gedankenexperiment.}
	\label{figure}
\end{figure*}

\subsection*{The plan}

The aim of this note is two-fold:
\begin{enumerate}
	\item Highlight that the original Wigner's friend paradox admits a classical explanation. That is, the contradicting predictions at the core of the gedankenexperiment arise in a simple non-contextual classical model, in which different agents have different (and conflicting) descriptions. Here, `contradicting predictions' refers to the fact that Wigner's and the friend's descriptions give different predictions for the possible outcomes of Wigner's entangled measurement.  While this will not come as a surprise in the quantum foundations community, we think it is worth to present it in some detail, since it shows that recent `supercharged' extensions of Wigner's friend \cite{brukner2017, frauchiger2018, pusey2016is, healey2018quantum,bong} are necessary to avoid simple classical explanations.
	\item Within such toy model, showcase that the ``paradoxes'' in the way quantum mechanics deals with macroscopic agents admit resolutions that are basically a reiteration of Jaynes' resolutions of thermodynamic ``paradoxes'', such as Gibb's paradox and Loschmidt's paradox (see Fig.~\ref{figure}). We submit that these connections between the foundations of quantum mechanics and thermodynamics, while they cannot be straightforwardly extended to arbitrary quantum settings, may nonetheless give suggestive hints for potential resolutions of these extremely challenging problems.
\end{enumerate}

We note that the possibility of such a classical model has been suggested previously in Refs.~\cite{frauchiger2018,cavalcanti2021view}. However, these works do not provide an explicit classical ontology and corresponding model, as we do in this work. We also stress that the toy model we present here does not refute the claims of recent extensions to the original Wigner's friend paradox \cite{brukner2017,frauchiger2018,pusey2016is,healey2018quantum,bong}, which feature multiple encapsulated friends. Indeed, a purely classical model like the one we present here is ruled out by the no-go theorems presented in these works. We expand on this issue in section \ref{sec:outlook}. We also note a number of interesting recent works \cite{w1,w2,w3,w4,w5,w6,cavalcanti2021view} from a fast-increasing literature on Wigner's friend-type experiments.

\section{Standard resolutions of the paradox}
Before presenting our toy model, we briefly review the standard resolutions of the original Wigner's friend paradox offered by the many-worlds, Bohmian and QBist interpretations of quantum theory, pointing out how each requires a profound revision of classical notions of reality.

\textbf{Many-worlds}--- This interpretation \cite{dewitt2015many} avoids the paradox of Wigner's friend by denying the objective existence of wavefunction collapse. Rather, all phenomena are seen as the consequence of a universal, Schr\"{o}dinger-evolving wavefunction. According to many-worlds, the description of Wigner is in fact the correct one and the friend and system really are in the entangled state $\ket{\phi^+}_{SF}=\frac{1}{\sqrt{2}} (\ket{0}_S \otimes \ket{0}_F  + \ket{1}_S \otimes \ket{1}_F )$.\footnote{More precisely, since there is no collapse in many-worlds, all three systems will be in an enormous entangled state involving all systems with which they have interacted in the past.} The two terms of this state correspond to two branches of the universal wavefunction; in each branch there exists a version of the friend observing one of the possible measurement outcomes. Thus, even the friend, finding herself in one of these branches with equal probability, understands that she is in fact entangled with another copy of herself, and would agree with Wigner's prediction for the entangled measurement. From an Everettian standpoint, it is less natural to see a paradox around Wigner in the first place. However the resolution comes at the price of an extremely nonclassical ontology based on the universal wavefunction. In contrast to this, here we highlight that the paradox can be dissipated even within a completely classical ontology. In this context, our resolution has many similarities with the solution of apparent paradoxes involving violations of the second law of thermodynamics \cite{jaynes1992gibbs}. 

\textbf{Bohmian mechanics}--- Bohmian mechanics \cite{durr2012quantum} is a deterministic and realist interpretation of quantum theory, in which the state of a system is given by a point $x$ in the configuration space of its constituent particles' positions, together with the usual Schr\"{o}dinger-evolving wavefunction (referred to as the `pilot wave'). All particles have well-defined positions at all times, although one requires that initial particle positions be uncertain and distributed as $\vert \psi (x,t=0)\vert^2$. The positions of the particles evolve deterministically in a manner governed by the pilot wave, in such a way that this uncertainty evolves as $\vert \psi(x,t)\vert^2$ as predicted by the Schr\"{o}dinger unitary evolution, thus reproducing the quantum predictions. 

In this interpretation, the friend sees a single, definite outcome that is determined from the initial particle positions. Due to the initial uncertainty in these positions, this outcome will be unknown to Wigner. His description corresponds to an incredibly complex many-particle state in which the positions of the particles constituting $F$ (her measurement device and, ultimately, her brain) have become correlated with the initial positions of the system, together with the unitarily evolving wavefunction (or pilot wave) that appears in the many-worlds interpretation. Like the many-worlds interpretation, in Bohmian mechanics there is no objective collapse of the wavefunction. The description of Wigner above is thus the correct one. One may however talk about an `effective collapse' in Bohmian mechanics, which can be applied if the macroscopic entanglement that has developed between the system and the measuring device does not play a role in later dynamics. That's the typical case, since entangling a system with its surroundings induces decoherence which, for all practical purposes, is irreversible. In the Wigner's friend experiment this assumption is clearly not justified, since the experiment is designed specifically to exploit this entanglement (see also Ref.~\cite{lazarovici2019quantum}). 

We note that although this resolution appears somewhat classical, due to the existence of definite particle positions, Bohmian mechanics still features highly nonclassical elements. Namely, the theory is both non-local and contextual and, just like in many-worlds, the universal wavefunction remains part of the ontology of the theory.

\textbf{QBism}--- QBism~\cite{fuchs2019qbism} is an interpretation in which quantum states, unitary evolution and measurement operators are all understood as personal judgments of individual agents, in a similar way to the subjective Bayesian interpretation of classical probability theory. As a result, the wavefunction does not encode objective facts about reality, but only subjective beliefs of an individual agent about the results of future experiences that he or she may have. For a QBist, in fact, even measurement outcomes are not objective features of the world, but they are only part of a single agent's private experience. This solves the disagreement between Wigner and his friend by rejecting the existence of observer-independent facts about measurement outcomes, which are accepted in the standard Copenhagen interpretation.

\section{Wigner friend's within a realist toy model}

\subsection{The toy model}

The toy model we use to analyse the experiment is inspired by (but not identical to) Spekkens' toy model for two qubits~\cite{spekkens2007evidence}. We imagine that each qubit system $X$ carries two labels, $\mathcal{X}$ and $\mathcal{X}'$, taking binary values $(i,j)$. $\mathcal{X}$ determines the outcome of a computational basis measurement $i=0,1$, while $\mathcal{X}'$ corresponds to some other internal degree of freedom. Just as in statistical mechanics, the state of the system is well-defined at any time but not necessarily known. The state is hence described by a probability distribution~$p^{ij}$:
\small
\begin{equation}
\label{eq:distribution}
\vec{p}_{\mathcal{X} \mathcal{X}'} = \sum_{i,j=0}^1 p^{ij}\; \vec{i}_{\mathcal{X}} \otimes \vec{j}_{\mathcal{X}'} = (p_{00},p_{01},p_{10},p_{11})^{\textrm{T}},
\end{equation}
\normalsize
where $\vec{0}_\mathcal{X}$, $\vec{1}_\mathcal{X}$ are canonical basis vectors $(1,0)^T$, $(0,1)^T$ on $\mathcal{X}$, and similarly for $\mathcal{X}'$ (we will drop $\otimes$ from now on for simplicity). Eq.~\eqref{eq:distribution} just means ``the state is $(i,j)$ with probability $p^{ij}$''. We postulate that, whenever $\ket{0}_X$ is prepared, $\mathcal{X}$ is set to $0$ while $\mathcal{X}'$ is uniformly randomly prepared in either $j=0$ or $j=1$. Similarly, whenever $\ket{1}_X$ is prepared  $\mathcal{X}$ is set to $1$ and $\mathcal{X}'$ is randomly initialised. Defining for convenience $\vec{u}_{\mathcal{X}'}:= \frac{1}{2} \vec{0}_{\mathcal{X}'} + \frac{1}{2} \vec{1}_{\mathcal{X}'}$, the correspondence between the quantum formalism and the toy model goes as
\begin{align}
\ket{0}_X \mapsto \vec{p}_{\mathcal{X} \mathcal{X}'} = \vec{0}_\mathcal{X} \vec{u}_{\mathcal{X}'}, \\
\ket{1}_X \mapsto \vec{p}_{\mathcal{X} \mathcal{X}'} = \vec{1}_\mathcal{X} \vec{u}_{\mathcal{X}'}.
\end{align}  
Furthermore, the Hadamard unitary on a system $X$ in quantum mechanics is described as a CNOT operation between $\mathcal{X}$ and $\mathcal{X}'$ controlled on $\mathcal{X}'$ in the toy model.
The preparation of product quantum states corresponds to product probability distributions\footnote{Note in passing that the PBR theorem~\cite{pusey2012reality} prevents models with this property to reproduce the whole of quantum mechanics.}, i.e.\ a product of qubit states on $X$ and $Y$ is associated to the probability distribution $\vec{p}_{\mathcal{X}\mathcal{X}'}\vec{p}_{\mathcal{Y} \mathcal{Y}'}$ in the toy model. 
Finally, a CNOT between $X$ and $Y$ (controlled on $X$) in quantum mechanics is described as two CNOTs in the model: one between $\mathcal{X}$ and $\mathcal{Y}$ (controlled on $\mathcal{X}$) and one between $\mathcal{X}'$ and $\mathcal{Y}'$ (controlled on $\mathcal{Y}'$).

\subsection{Wigner's friend description of the experiment}

The initial state of $S$ is $\ket{0}_S$, described as  $\vec{0}_\mathcal{S} \vec{u}_{\mathcal{S}'}$ in the toy model. After the Hadamard unitary the quantum state is $\ket{+}_S$ which, according to the above rules, is represented as a correlated distribution
\begin{equation}
\label{eq:plus}
\ket{+}_S \mapsto \frac{1}{2} \vec{0}_\mathcal{S} \vec{0}_{\mathcal{S}'} + \frac{1}{2} \vec{1}_\mathcal{S} \vec{1}_{\mathcal{S}'}.
\end{equation}

If the friend performs a computational basis measurement, half of the times she sees $\mathcal{S}$ in $0$ (so her memory $\mathcal{F}$ records $0$) and the other half she sees $\mathcal{S}$ in $1$ (so her memory $\mathcal{F}$ records $1$). Crucially, her level of description does not involve a dynamical account of the measurement process, which may introduce correlations (unknown to her) between herself and the system. Hence her best guess is to describe the systems after the measurement by the most unbiased distribution compatible with her knowledge. This is the maximum entropy distribution:\footnote{The original Spekken's original toy model~\cite{spekkens2007evidence} instead postulates a disturbance that re-randomises the primed degrees of freedom. The ignorance interpretation we adopt, however, plays an important role in the toy model explanation of the original Wigner's friend paradox.} 
\begin{equation}
\label{eq:friend1}
\textrm{If  } \, (\mathcal{S}=0, \mathcal{F}=0): \quad \vec{0}_\mathcal{S}  \vec{u}_{\mathcal{S}'} \vec{0}_\mathcal{F} \vec{u}_{\mathcal{F}'}. 
\end{equation}
\begin{equation}
\label{eq:friend2}
\textrm{If  } \, (\mathcal{S}=1, \mathcal{F}=1): \quad \vec{1}_\mathcal{S}  \vec{u}_{\mathcal{S}'} \vec{1}_\mathcal{F} \vec{u}_{\mathcal{F}'}. 
\end{equation}

Note that these are the toy model representations of $\ket{0}_S \otimes \ket{0}_{F}$ and $\ket{1}_S \otimes \ket{1}_{F}$. This procedure corresponds, in the quantum scenario, to the collapse rule telling us that, after the measurement, the state is one of these two states with equal probability. We will see that Eqs.~\eqref{eq:friend1}-\eqref{eq:friend2} can in fact be obtained as a coarse-graining of Wigner's description that ignores the $\S\F/\S'\F'$ correlations generated by the measurement dynamics. Hence in the toy model irreversibility is a consequence of the friend's coarse-graining.

At first sight the friend's coarse-graining procedure leading to Eqs.~\eqref{eq:friend1}-\eqref{eq:friend2} may appear puzzling, since we are simply dealing with binary degrees of freedom whose evolution can be easily tracked. To avoid confusion one has to keep in mind that the toy model tries to capture the essence of a much more complex situation. In reality we should think of $S$ as interacting with a system $F$ composed of an Avogadro number of constituents -- so large as to make a complete dynamical account of the measurement process practically unfeasible for human-scale agents. Just like agents in thermodynamics have to resort to a coarse grained description of a gas in a box in terms of certain macroscopic variables such as pressure, volume etc., Wigner's friend $F$ needs a coarse-grained description where the measurement device is simply described in terms of the degrees of freedom displaying the measurement outcome. Pushing the analogy, at this ``macroscopic'' level of description the gas in a box appears to irreversibly approach equilibrium and the measurement device appears to irreversibly display an outcome. But the underlying dynamics in both cases is fundamentally reversible, so we can theoretically conceive of extremely powerful agents that can reverse it. In thermodynamics such agents are called Maxwell's demons \cite{maruyama2009colloquium}. In quantum foundations, they go by the less exciting name of Wigner.

\subsection{Wigner's description of the experiment}

Wigner describes the measurement processes dynamically, as an interaction between $S$ and $F$. Differently from $F$, however, he does not know the measurement outcome. The state just before the $SF$ interaction is
\begin{equation}
\left(\frac{1}{2} \vec{0}_\S \vec{0}_{\S'} + \frac{1}{2} \vec{1}_\S \vec{1}_{\S'}\right)\vec{0}_\F \vec{u}_{\F'},
\end{equation}
where, to keep the model simple, we took $\F = 0$ before the interaction to signify the  ``ready'' state.\footnote{We could have otherwise added a third, ``ready'' state for $F$, but this is unnecessary.} The $SF$ interaction is described by a CNOT in quantum mechanics, and as two CNOTs in the toy model: a $\S\F$ CNOT and $\F'\S'$ CNOT, where the first label indicates the control system. After the interaction we get
\small
\begin{align}
\frac{1}{4}\left( \vec{0}_\S \vec{0}_{\S'}  \vec{0}_\F \vec{0}_{\F'}  + \vec{0}_\S \vec{1}_{\S'}  \vec{0}_\F \vec{1}_{\F'} 
 +\vec{1}_\S \vec{1}_{\S'}  \vec{1}_\F \vec{0}_{\F'}  + \vec{1}_\S \vec{0}_{\S'}  \vec{1}_\F \vec{1}_{\F'}
\right).
\label{eq:finalwigner}
\end{align}
\normalsize
The friend's ``experience'' of being in a superposition is nothing weird -- in the toy model the superposition corresponds Wigner's more refined description of the same state of affair. She really always is in a well-defined state, independently of what the various agents know about it. 

Note that Wigner's description \emph{cannot} be recovered from the friend's simply by averaging Eq.~\eqref{eq:friend1}-\eqref{eq:friend2} to account for Wigner's lack of knowledge of her measurement outcome.
In fact such averaging gives
\begin{equation}
\label{eq:friendaverage}
\frac{1}{2} \left(\vec{0}_S \vec{0}_\F + \vec{1}_\S \vec{1}_\F \right) \vec{u}_{\S'}  \vec{u}_{\F'}, 
\end{equation}
which excludes $\S\F/\S'\F'$ correlations. This is not surprising, since such correlations are beyond the friend's level of description. That is the toy model counterpart to the fact that in quantum theory the state is entangled only in Wigner's description. 

However Wigner's and the friend's descriptions of the marginals $\S\S'$ or $\F\F'$ do coincide, once one takes into account Wigner's ignorance of the measurement outcome. The reason the friend can get away with her coarse-grained description is that she will not be found at fault as long as the abovementioned $\S\F/\S'\F'$ correlations do not come into play in later dynamics. Given the size of $F$, not speaking of the rest of the environment typically involved, one expects that such correlations will play no role unless some Maxwell's demon-like agent comes into play. Note the analogy with standard thermodynamic descriptions of the interaction between a system and a large environment, in which system-environment correlations are neglected~\cite{breuer2002theory}. Situations in which such correlations come back into play are called \emph{information backflows} and they typically become increasingly unlikely as the size of the environment grows. In fact, it has been suggested \cite{milazzo2019role} that the absence of such backflows are a relevant feature in the quantum-classical transition via quantum Darwinism \cite{zurek2009quantum}. We can clearly see the classical counterpart within the toy model.

What if Wigner measures the state of the friend? If he does not describe its own measurement process dynamically (again, say due to practical limitations), then everything he knows is that he sees either $\F = 0$ or $\F = 1$, while $\F'$ \emph{after the measurement} is completely unknown to him (in fact, an agent even more powerful than Wigner would describe this by means of correlations between $\F'$ and $\mathcal{W}'$). Leaving out Wigner's description of himself, if he performs Baysian update on Eq.~\eqref{eq:finalwigner} and sets $\F'$ to be uniform, his description will now coincide with that of his friend: he gets either Eq.~\eqref{eq:friend1} or Eq.~\eqref{eq:friend2}, depending on the outcome. In summary, if Wigner is not able or willing to describe his own measurement dynamics, he can gain access to his friend's measurement outcome only at the price of renouncing to his demon-status. Specifically, Wigner loses the information he needed to reverse the friend's measurement dynamics. In quantum mechanics we say that, by measuring, Wigner has collapsed  the previously entangled state.\footnote{Note that an agent able to access the measurement result while accounting for the full dynamics is logically consistent with the spirit of the toy model, but they would be a supra-quantum agent.} 

\subsection{A realist's resolution of the puzzle}

We now use this toy model to discuss the controversies engendered by the original Wigner's friend experiment and their compatibility with the notion of objective reality.

Consider Wigner's entangled measurement, whose quantum description in practice looks as follows: he has to perform a CNOT between $S$ and $F$ (controlled on S), followed by a Hadamard on $S$ and a computational basis measurement on the two, with the $4$ outcomes $00, 01, 10, 11$ corresponding to the $4$ outcomes $\phi^+$, $\phi^-$, $\psi^+$, $\psi^-$. Unsurprisingly, what this does in the toy model is just to reverse the dynamics all the way back to the initial state
\begin{equation}\label{initial}
\vec{0}_\S \vec{u}_{\S'}  \vec{0}_\F \vec{u}_{\F'}.
\end{equation}
The friend's memory of the fact her measurement has ever happened has been erased by Wigner. As far as the friend is concerned, she's still in the ``ready'' state and of course she will agree with Wigner that his computational basis measurement will return the outcome 00 (which corresponds to outcome $\phi^+$ of the entangled measurement). \emph{Wigner's creation of a measurement outcome involves time-reversing the dynamics that created his friend's outcome.} 

Nothing mysterious is happening here. If the friend is told that the superagent Wigner is about to perform a measurement involving both $S$ and $F$, she is aware that her level of description may be insufficient to correctly predict what is going to happen. The friend's description is still rational -- from a maximum entropy principle perspective it is the best predictions she could make within her level of description and given the evidence she had.

Note that, as in the original Wigner's friend `paradox', if Wigner adopts the friend's description for either of her measurement outcomes, he will not predict that the outcome corresponding $\phi^+$ is obtained with certainty, but rather predict that both $\phi^+$ and $\phi^-$ are equally likely to occur. Let's see this. Wigner's entangled measurement is described quantum mechanically as a CNOT between $S$ and $F$, followed by a Hadamard on $S$ and a computational basis measurement on $S$, $F$, with outcomes $00$, $10$, $01$, $11$ corresponding to $\phi^+$, $\phi^-$, $\psi^+$, $\psi^-$ respectively. So in the toy model the measurement is described by three CNOT operations (in order, $\S \F$, $\F' \S'$, $\S' \S$, with the first label the control) followed by a reading out of $\S$, $\F$ and re-randomization of $\S'$, $\F'$. Applying the $3$ CNOTs to the friend's description of the states after her measurement (Eq.~ \eqref{eq:friend1}, \eqref{eq:friend2}) we get  %
\small
\begin{align}
\frac{1}{4}(\vec{0}_\S \vec{0}_{\S'}  \vec{0}_\F \vec{0}_{\F'}  +\vec{1}_\S \vec{1}_{\S'}  \vec{0}_\F \vec{0}_{\F'}  +\vec{1}_\S \vec{1}_{\S'}  \vec{0}_\F \vec{1}_{\F'}  +\vec{0}_\S \vec{0}_{\S'}  \vec{0}_\F \vec{1}_{\F'}  )
\end{align}
\normalsize
for $\S=0$, $\F=0$ and 
\small
\begin{align}
\frac{1}{4}(\vec{1}_\S \vec{0}_{\S'}  \vec{0}_\F \vec{0}_{\F'}  +\vec{0}_\S \vec{1}_{\S'}  \vec{0}_\F \vec{0}_{\F'}  +\vec{0}_\S \vec{1}_{\S'}  \vec{0}_\F \vec{1}_{\F'}  +\vec{1}_\S \vec{0}_{\S'}  \vec{0}_\F \vec{1}_{\F'}  )
\end{align}
\normalsize
for $\S=1$, $\F=1$. A reading out of $\S$, $\F$ now returns the outcomes $00$ and $10$ with equal probability in either case, which corresponds to the outcomes $\phi^+$, $\phi^-$ in the quantum experiment. Thus, the fundamental disagreement is recovered in the toy model.

To pursue the thermodynamic analogy further,  a macroscopic agent describing a gas in a box knows that their predictions are invalid if a more powerful agent (Maxwell's original demon being an extreme example) comes by with control over extra thermodynamic variables. The superagent  can conjure a violation of the second law of thermodynamics, e.g. extracting work from (what the less powerful agent describes as) an equilibrium thermal state \cite{jaynes1965gibbs, jaynes1992gibbs}.\footnote{``Equilibrium thermal state'' is necessarily an agent-dependent notion, since there's no ultimate ``true'' thermal state. After all, in classical physics the state of the system really is a point in phase space. For an agent like Maxwell's original demon, who knows positions and momenta of all particles, there's no statistical mechanics~\cite{jaynes1965gibbs}.} None of this invalidates the theory according to which the gas is made of particles with definite positions and momenta determining the outcome of any experiment. Similarly, as the toy model shows, Wigner's and the friend contrasting predictions peacefully coexist with an underlying, objective state of the world evolving reversibly.

As a general comment note that in realist theories measurement outcomes do not have a fundamentally privileged role compared to any other degree of freedom. That is, in any such (reversible) theory, 1. outcomes are the result of dynamical processes and are encoded in objective states of the world, and 2. their creation can be undone by a sufficiently powerful ``Wigner demon''. In quantum mechanics, the different agents' levels of description correspond to putting the Heisenberg cut in different places -- but quite clearly there's no special place where the cut suddenly becomes objective. Within the toy model, the question: Where does the collapse \emph{actually} happen? is meaningless. It is set at the level above which we do not wish to or can track the full dynamics. 

Furthermore, there is another aspect in which the analogy with thermodynamics is striking. As we mentioned, it is often harmless to talk about \emph{the} second law without making any reference to its anthropomorphic origin, because \emph{in practice} there is an essentially unique, reasonable way to define the thermodynamic variables. At the foundational level, however, forgetting this anthropomorphic origin \cite{jaynes1965gibbs} leads to many problems, including: 
\begin{enumerate}
	\item \emph{Loschmidt's paradox}: how is macroscopic irreversibility compatible with a microscopically reversible theory?
	\item \emph{Gibbs' paradox}: a more informed agent can trick a lesser informed one by undoing a seemingly irreversible process -- extracting work from a thermal state, pumping heat from a cold to a hot body etc. This is in violation of ``the'' second law of thermodynamics as seen by the less powerful agent \cite{jaynes1992gibbs}. 
\end{enumerate}

Similarly, we can conveniently talk of \emph{the} measurement collapse, because \emph{in practice} there is an essentially unique, reasonable way to define it. The toy model analogously suggests that forgetting the anthropomorphic origin of the cut leads to similar sounding problems:
\begin{enumerate}
	\item How is irreversible collapse compatible with reversible unitary dynamics?
	\item Wigner tricks his friend by undoing a seemingly irreversible process. This is in violation of the friend's predictions.
\end{enumerate}

Not only these problems sound similar to the thermodynamic ones -- within the toy model they \emph{are} essentially the same. 

\section{Outlook}\label{sec:outlook}

Of course the toy model can only reproduce a subset of quantum theory, so these issues evaporate only within it. All the same, these problems are often presented as puzzling issues making only reference to the original Wigner's friend experiment or similar settings. The existence of the toy model then forces us to answer the question: \emph{what operational aspects of these problems cannot be explained within a realist and essentially classical mindset?} By `classical' here we allow for theories like Spekken's toy model, where there are intrinsic limitations to the agents' ability to specify the initial state, but we do not allow stronger forms of nonclassicality, such as nonlocality and contextuality \cite{jennings2016no}. 

One could look at the sequence of closely-related extensions to the original Wigner's friend thought experiment, first introduced by Brukner \cite{brukner2017}, and followed by Frauchiner \& Renner \cite{frauchiger2018}, Masanes \& Pusey~\cite{pusey2016is, healey2018quantum}, and Bong et. al. \cite{bong}. These extensions are ``mashups'' of the original Wigner's friend and Bell scenarios, in which a pair of encapsulated observers at different locations share and measure an entangled state, violating a Bell inequality in the process. Our toy model formally proves that these extensions are needed to extract a behaviour that defies classical explanations. Since these extensions contain nonlocality, a local and non-contextual model in the spirit of our toy model is not possible. The next question is then whether the recent mashups contain a result that is stronger and independent of the no-go results due to Bell \cite{bell} and Kochen-Specker \cite{ks}. Indeed, in Ref.~\cite{bong} it is shown that such experiments do imply stronger constraints on the kind of hidden variable theories reproducing quantum theory. In particular, in \cite{bong} it is shown that any hidden variable theory satisfying conditions called `local friendliness' are ruled out by quantum theory. Since our model satisfies local friendliness it cannot be extended to account for these scenarios without introducing non-classical elements. Nevertheless, our work may provide a route to understanding such scenarios in a model that, although not satisfying local friendliness, nevertheless retains certain elements of classicality. That should be taken part of a long-term, extremely challenging research program that savages the spirit of local realism while including the highly nonclassical features implied by all the recent no-go theorems. It is our opinion that this requires a change of framework, going beyond the standard notion of ontological models. For an extended presentation of this point of view, see Ref.~\cite{schmid2020unscrambling}.

While the thermodynamic analogies discussed within the toy model can be expected to be only partially applicable to quantum mechanics as a whole, they give some suggestive hints for the realist camp. There are strong arguments supporting that classical thermodynamics can be seen as a theory of best inferences from partial information~\cite{jaynes1965gibbs, jaynes1992gibbs}. At the same time, in classical mechanics, one can go quite some way in the attempt to explain the physical circumstances that make thermodynamics possible and useful. That is, one can derive that, in the late-time physics, confined subsystems with conserved quantities and physical rods and clocks emerge typically from the structure of the solutions of Newton's gravitational law \cite{barbour2014identification}. To keep with the analogy, then, one may hope that the quantum structures emerge as a theory of best inferences within subsystems in the late-time physics from some model of the whole universe through a \emph{conceptually} similar mechanism. At least, we believe that would be a program worth attempting.
 
\bigskip

{\bf Acknowledgments.} We are thankful to Terry Rudolph, Gabriel Senno, John Selby, Chris Fuchs and David Schmid for helpful comments on an earlier draft. JB acknowledges funding from the Spanish MINECO (Severo Ochoa SEV-2015-0522) and Fundacio Cellex, Mir-Puig, Generalitat de
Catalunya (SGR 1381 and CERCA Programme), and the AXAChair in Quantum Information Science. ML acknowledges funding from the EU Marie Sklodowska-Curie individual Fellowships (H2020-MSCA-IF-2017, GA794842), Spanish MINECO (Severo Ochoa SEV-2015-0522, project QIBEQI FIS2016-80773-P), Fundacio Cellex and Generalitat de Catalunya (CERCA Programme and SGR 875) and ERC Grant EQEC No. 682726.

\bibliography{Bibliography}

\end{document}